\documentclass[aps,prl,twocolumn,showpacs,superscriptaddress,groupedaddress]{revtex4}  

\usepackage{graphicx,amssymb,amsfonts,amsmath,amssymb,amscd,amstext}
\pdfoutput=1

\def\be{\begin{equation}}
\def\ee{\end{equation}}

\def\tr{\operatorname{tr}}
\def\calo{{\mathcal O}}
\def\i{{\rm i}}

\begin{document}

\title{Universal Thermal Corrections to Single Interval Entanglement Entropy for Two Dimensional Conformal Field Theories}

\author{John Cardy}
\affiliation{Oxford University, Rudolf Peierls Centre for Theoretical Physics \\
1 Keble Road, Oxford, OX1 3NP, United Kingdom
}
\affiliation{All Souls College, Oxford}

\author{Christopher P. Herzog}
\affiliation{C.~N.~Yang Institute for Theoretical Physics, 
Department of Physics and Astronomy \\
Stony Brook University, Stony Brook, NY  11794
}
\date{March 3, 2014}

\begin{abstract}
We consider single interval R\'enyi and entanglement entropies for a two dimensional conformal field theory on a circle at nonzero temperature.  Assuming that the finite size of the system introduces a unique ground state with a nonzero mass gap, we calculate the leading corrections to the R\'enyi and entanglement entropy in a low temperature expansion.  These corrections have a universal form for any two dimensional conformal field theory that depends only on the size of the mass gap
and its degeneracy. We analyze the limits where the size of the interval becomes small and where it becomes close to the size of the spatial circle.
\end{abstract}

\maketitle

The idea of entanglement,  that a local measurement on a quantum system may instantaneously affect the outcome of local measurements far away, is a central concept in quantum mechanics.  As such, it underlies the study of quantum information, communication and computation.  However, entanglement also plays an increasing role in other areas of physics.  To name three, measures of entanglement may detect exotic phase transitions in many-body systems  lacking a local order parameter \cite{Osborne:2002zz,Vidal:2002rm};  such measures order quantum field theories under renormalization group flow \cite{Casini:2006es,Casini:2012ei}; entanglement is a key concept in the black hole information paradox (see e.g. 
\cite{Bombelli:1986,Srednicki:1993im}).  

Important measures of entanglement for a many-body system in its ground state are the R\'enyi and entanglement entropies.
To define these quantities, we first partition the Hilbert space into pieces $A$ and complement $\bar A = B$.  
Typically (and hereafter in this letter) $A$ and $B$ correspond to spatial regions.  
Not all quantum systems may allow for such a partition.
The reduced density matrix is defined as a partial trace of the full density matrix $\rho$ 
over the degrees of freedom in $B$: 
\be
\rho_A \equiv \tr_B \rho \ .
\ee
The entanglement entropy is then the von Neumann entropy of the reduced density matrix:
\be
S_E \equiv - \tr \rho_A \log \rho_A \ .
\label{EEdef}
\ee
The R\'enyi entropies are
\be
S_n \equiv \frac{1}{1-n} \log \tr (\rho_A)^n \ ,
\label{Renyidef}
\ee
where $S_E = \lim_{n \to 1} S_n$.  

The entanglement and R\'enyi entropies are well defined for excited and thermal states as well.  One simply starts with the relevant excited state or thermal density matrix instead of the ground state density matrix
$\rho = | 0 \rangle \langle 0 |$.
For example, for thermal states, one would start with
\be
\rho = \frac{e^{-\beta H}}{\tr( e^{-\beta H})} \ ,
\label{thermalrho}
\ee
where $\beta$ is the inverse temperature and $H$ is the Hamiltonian.
However, it is well known that for mixed states, the entanglement entropy is no longer a good measure of quantum entanglement.
% Maybe omit any reference here? (see e.g.\ \cite{Calabrese:2009qy} [[ better reference ? ]]).  
The entanglement entropy is contaminated by the thermal entropy of region $A$ and in fact in the high temperature limit becomes dominated by it.    

It is of course never possible to take a strict zero temperature limit of real world systems.  
If we are to use entanglement entropy to measure quantum entanglement, then we need to know how to subtract off nonzero temperature contributions to $S_E$ and $S_n$.  Little has been said about these corrections in the literature thus far.
Ref.\ \cite{Herzog:2012bw} studied entanglement entropy for a massive relativistic scalar field in 1+1 dimensions.
The authors provided numerical evidence that the corrections scale as $e^{-\beta m}$ in the limit $\beta m \gg 1$ 
where $m$ is the mass of the scalar field.  They further conjectured that for any many-body system with a mass gap $m_{\rm gap}$,
such corrections should scale as $e^{-\beta m_{\rm gap}}$ when $\beta m_{\rm gap} \gg 1$.  This conjecture more or less immediately follows from writing (\ref{thermalrho}) as a Boltzmann sum over states.  By assumption, the first excited state will come with contribution $e^{-\beta m_{\rm gap}}$ and give the dominant correction.  Some possible but probably rare exceptions involve cases where for some reason the first excited state contribution to the entanglement entropy vanishes or where the degeneracy of states with energy $E = m_{\rm gap}$ itself has an exponential dependence on temperature.
  
While knowledge of the $e^{-\beta m_{\rm gap}}$ dependence is a good starting point, it would be even better to know the coefficient.  In this letter we calculate the coefficient for 1+1 dimensional conformal field theories (CFTs) on a circle of circumference $L$ in the case where $A$ consists of a single interval of length $\ell$.  
The answer is
\begin{eqnarray}
\label{dSncorr}
\delta S_n &=& \frac{g}{1-n} \left[\frac{1}{n^{2\Delta-1}} \frac{ \sin^{2\Delta} \left( \frac{ \pi \ell}{L} \right) }{\sin^{2\Delta} \left( \frac{ \pi  \ell}{n L} \right)} -n\right] 
e^{-2 \pi \Delta \beta / L} +\nonumber \\
&& o(e^{-2 \pi \Delta \beta / L}) 
\  , \\
\label{dSEcorr}
\delta S_E &=& 2g \Delta \left[1 - \frac{\pi \ell}{L} \cot \left( \frac{ \pi \ell}{L} \right) \right] e^{-2 \pi \Delta \beta/ L} \nonumber \\
&& + o(e^{-2 \pi \Delta \beta / L})\ ,
\end{eqnarray}
where $\Delta$ is the smallest scaling dimension among the set of operators including the stress tensor and all primaries not equal to the identity and $g$ is their degeneracy.
%and $\ell$ is the length of $A$.  
In order for this result to hold, the CFT needs to have a unique ground state separated from the first excited state by a nonzero mass gap (induced by the finite volume of the system).  
The correction (\ref{dSEcorr})  was obtained from the $n \to 1$ limit of (\ref{dSncorr}).
Given the form of these corrections, another possible use for them is the determination of $\Delta$ and $g$. 

From the literature \cite{Azeyanagi:2007bj,Herzog:2013py,McIntyre:2004xs,Barrella:2013wja}, it is possible to deduce the correction (\ref{dSncorr}) in two limiting cases -- for free theories and for theories with AdS/CFT duals and large central charge.
Using bosonization, ref.\ \cite{Azeyanagi:2007bj,Herzog:2013py} computed these corrections for massless free Dirac fermions.  
In this case, free fermionic operators with scaling dimension $\Delta=1/2$ and $g=4$ yield the appropriate correction.
	Ref.\ \cite{Barrella:2013wja} considered CFTs with large central charge $c$ and gravitational duals.  In this case, the entanglement and R\'enyi entropies can be written as an asymptotic series in $c$ where the leading term is $\calo(c)$.  
The authors computed the $\calo(1)$ corrections to the entanglement and R\'enyi entropies due to the stress tensor $T(z)$, which has $\Delta=2$ and $g=2$.  Although the authors do not write explicit formulae, it is trivial to extend their computation to operators of general conformal dimension.  The computation involves a low temperature expansion of a Selberg-like zeta function.  In all cases, their expression would match (\ref{dSncorr}).  

For free CFTs more generally, the important observation is that this same Selberg-like zeta function appears in ref.\ \cite{McIntyre:2004xs} where the authors compute the determinant of the Laplacian for a $bc$-system on an arbitrary genus Riemann surface.  From ref.\ \cite{McIntyre:2004xs}, one can then deduce that the same correction (\ref{dSncorr}) holds for free CFTs \footnote{%
The authors of ref.\ \cite{McIntyre:2004xs} demonstrate their formula only for a $bc$-system with integer scaling dimension, but it is likely to hold more generally \cite{Takhtajan}.
}.
A check of these statements was performed explicitly for a free complex boson and free fermions in ref.\ \cite{Datta:2013hba}.
%[[ I think for free theories, we can relax the requirement of a unique ground state. ]]

In this letter, we will show by explicit computation that the corrections (\ref{dSncorr}) and (\ref{dSEcorr}) hold for a general 1+1 dimensional CFT.  We then check that the corrections have expected behavior in the limits $\ell \ll L$ and $\ell \to L$.

\vskip 0.1in

\noindent
{\it Universal Thermal Corrections:}
Starting from the thermal density matrix (\ref{thermalrho}), 
we compute the first few terms in a low temperature expansion of the R\'enyi entropies. 
Introducing a complete set of states, we can write the thermal density matrix as a Boltzmann sum over states:
\be
\rho = \frac{1}{\tr(e^{-\beta H})} \sum_{|\phi \rangle} |\phi \rangle \langle \phi | \, e^{-\beta E_{\phi} } \ .
\ee
For a CFT on a cylinder of circumference $L$, coordinatized by $w= x - \i t $, the Hamiltonian is 
\be
H = \left( \frac{2\pi}{L} \right) \left(L_0 + \tilde L_0 - \frac{c}{12}\right) \ ,
\ee
where $L_0$ and $\tilde L_0$ are the zeroth level left and right moving Virasoro generators and $c$ is the central charge.
We will assume in what follows that placing the CFT on a cylinder produces a unique ground state and 
gaps the theory.
Let $\psi(w) \neq 1$ be a Virasoro primary operator and $|\psi \rangle = \lim_{t \to -\infty} \psi(w) |0 \rangle$ the corresponding state where $L_0 |\psi \rangle = h |\psi \rangle$, $\tilde L_0 | \psi \rangle = \tilde h | \psi \rangle$ and $\Delta = h + \tilde h$.
As the $|\psi \rangle$ are lowest weight states in the CFT, it follows -- with one important exception -- that the smallest nonzero $E_\phi$ must correspond to a primary operator $\psi(w)$ and moreover that $E_\psi = \frac{2 \pi}{L} \left( \Delta - \frac{c}{12} \right)$.  
The one important exception is when $\Delta > 2$ and two descendants of the identity operator, the stress tensor $T(w)$ and its conjugate $\tilde T(\bar w)$, give the dominant correction.  (The identity operator itself can be thought of as yielding the leading zero temperature contribution to the R\'enyi entropies.)
We are assuming $\Delta > 0$, which may not be true for non-unitary CFTs, and no gravitational anomaly, i.e.\ $c_L = c_R$.

 We divide up the spatial circle into regions $A$ and complement $B$ and compute the reduced density matrix $\rho_A$.
Let us consider a low temperature expansion where
\be
\label{rhoexpand}
\rho =\frac{\left( |0\rangle \langle 0 | + | \psi \rangle \langle \psi | e^{-2 \pi \Delta \beta /L} + \cdots \right)}
{1 +  e^{-2 \pi \Delta \beta / L} + \cdots }
 \ .
\ee
It follows then that
\begin{eqnarray}
\tr (\rho_A)^n &=& \frac
{\tr \left[ \tr_B \left( |0\rangle \langle 0 | + | \psi \rangle \langle \psi | e^{-2 \pi \Delta \beta /L} + \cdots \right) \right]^n}
{\left(1 +  e^{-2 \pi \Delta \beta / L} + \cdots\right)^n } \\
&=&
\tr \left( \tr_B |0 \rangle \langle 0 | \right)^n  \Biggl[ 1 +  \nonumber \\
&& \nonumber + \left( \frac{ \tr \left[ \tr_B | \psi \rangle \langle \psi | \left(\tr_B |0 \rangle \langle 0 | \right)^{n-1} \right] }{\tr \left( \tr_B |0 \rangle \langle 0 | \right)^n}- 1 \right) n e^{-2 \pi \Delta \beta/ L }  \\
&& + \ldots \Biggr]
\end{eqnarray}
The first term gives the zero temperature contribution to the R\'enyi entropy for a finite system
%\cite{Vidal:2002rm,Holzhey:1994we,Calabrese:2004eu}
\cite{Calabrese:2004eu}
\be
\log \tr (\tr_B |0 \rangle \langle 0 | )^n = c \frac{1-n^2}{6n} \log \left[ \frac{L}{\pi} \sin \left( \frac{ \pi \ell}{L} \right) \right] + \ldots
\ee
where $\ell$ is the length of $A$.

Up to a normalization issue we will come to shortly, the expression
\be
\label{twopt}
\frac{ \tr \left[ \tr_B | \psi \rangle \langle \psi | \left(\tr_B |0 \rangle \langle 0 | \right)^{n-1} \right] }{\tr \left( \tr_B |0 \rangle \langle 0 | \right)^n}
\ee
is a two-point function of the operator $\psi(w)$.  By the state operator correspondence, we can interpret the state $|\psi \rangle$ as  $\lim_{t \to -\infty} \psi(x,t) |0 \rangle$ and similarly $\langle \psi | = \lim_{t \to \infty} \langle 0 | \psi(x,t)$.  In path integral language, the trace over an $n$th power of the reduced density matrix becomes a partition function on an $n$-sheeted copy of the cylinder, branched over the interval $A$.
The expression (\ref{twopt}) is then the two-point function $\langle \psi(w_2, \bar w_2) \psi(w_1, \bar w_1) \rangle$ on this multisheeted cylinder for a particular choice of $w_1$ and $w_2$.

By the Riemann-Hurwitz theorem, this multisheeted cylinder has genus zero.  There exists then a uniformizing map which takes the multisheeted cylinder to the plane, and on the plane, the two-point function (\ref{twopt}) can be computed trivially.
A uniformizing map from $n$ copies of a cylinder of circumference $L$ to the plane is
\be
\zeta^{(n)} = \left( \frac{e^{2 \pi \i w/L} -  e^{\i \theta_2}}{e^{2 \pi i w/L} - e^{\i \theta_1}} \right)^{1/n}  \ .
\ee
We choose $\theta_1$ and $\theta_2$ such that the map is branched over the interval $A$, which 
has endpoints on the cylinder such that 
$
\theta_2 - \theta_1 = \frac{2 \pi \ell}{ L}
$.  
Under this map, an operator inserted on the $j$th cylinder at $t = -\infty$ is located on the $\zeta^{(n)}$ plane
at $\zeta^{(n)}_{-\infty} = e^{\i(\theta_2 - \theta_1)/n + 2 \pi \i j/ n}$, while an operator inserted at $t = \infty$ is located at
$\zeta^{(n)}_\infty = e^{2 \pi \i j/n}$.  

\begin{figure}
 \begin{center}
  \includegraphics[width=3.375in]{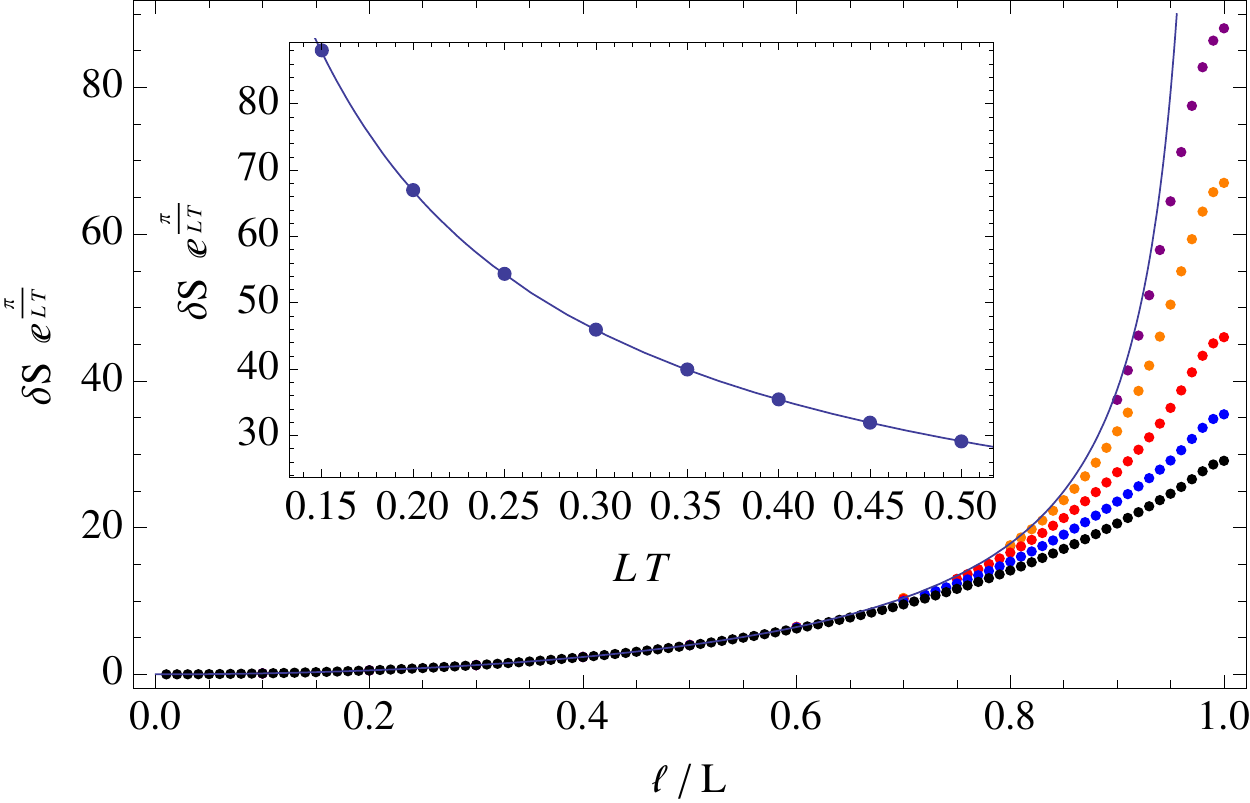}
 %b)
 %\includegraphics[width=2.8in]{thermal}
   \end{center}
  \caption{
  Thermal corrections to entanglement entropy $\delta S_E = S_E(T)-S_E(0)$ plotted against $\ell/L$ for a free, massless 1+1 dimensional fermion.  The points are numerically determined from a lattice model of size $N=100$ grid points using the method described in \cite{Herzog:2013py,EislerPeschel}. $L$ is the size of the circle and $\ell$ of the interval. From top to bottom, the points correspond to $LT = 0.15, 0.2, 0.3, 0.4, 0.5$.  The solid curve is the prediction (\ref{dSEcorr}). Inset: 
$\delta S_E$ Plotted against $LT$ for $\ell = L$. The curve is the thermal entropy correction $4 ( 1 + \pi / LT)$. 
 }
\label{EEplot}
 \end{figure}

The correlation function on the $\zeta$-plane 
between two quasi-primaries inserted at points $\zeta^{(n)}_2$ and $\zeta^{(n)}_1$ is
\begin{eqnarray}
\label{standardform}
 \left \langle\psi\left(\zeta^{(n)}_2,\bar \zeta^{(n)}_2 \right) \psi \left(\zeta^{(n)}_1, \bar \zeta^{(n)}_1 \right) \right \rangle =
   \frac{1}{\left( \zeta^{(n)}_{21}  \right)^{2h} \left( \bar \zeta^{(n)}_{21}  \right)^{2\tilde h}}  
\end{eqnarray}
using the standard CFT normalization for two point functions.
%(We restrict to chiral primaries in this section for notational ease, but it is straightforward to consider primary operators instead.) 
Mapping back to the $n$-sheeted cover of the $w$-plane, we find that
\begin{eqnarray}
\left \langle \psi(w_2, \bar w_2) \psi(w_1, \bar w_1) \right \rangle_n = 
\frac{ \left( \frac{d \zeta^{(n)}_1}{dw_1} \frac{d \zeta^{(n)}_2}{dw_2} \right)^h}
{ \left(\zeta^{(n)}_{21}  \right)^{2h}}
\frac{ \left( \frac{d \bar \zeta^{(n)}_1}{d \bar w_1}\frac{d \bar \zeta^{(n)}_2}{d \bar w_2} \right)^{\tilde h} }
{  \left(\bar \zeta^{(n)}_{21} \right)^{2\tilde h}} \ 
\label{almost}
\end{eqnarray}
where $\psi(w_1, \bar w_1)$ and $\psi(w_2, \bar w_2)$ are inserted on just one of the $n$-sheets.
We would like to work with a normalization such that $\langle \psi | \psi \rangle = 1$ on the original $w$-cylinder.  To that end, we need to divide (\ref{almost}) by $\langle \psi(w_2, \bar w_2) \psi(w_1, \bar w_1) \rangle_1$.
%\be
%\langle \psi(w_2) \psi(w_1) \rangle_1 = \left( \frac{d \zeta^{(1)}_1}{dw_1} \right)^h \left( \frac{d \zeta^{(1)}_2}{dw_2} \right)^h (\zeta^{(1)}_2 - \zeta^{(1)}_1)^{-2h} \ .
%\label{norm}
%\ee
The following identity is useful
\be
\frac{ d \zeta^{(n)} / d w}{d \zeta^{(1)} / dw} = \frac{1}{n} \frac{\zeta^{(n)}}{\zeta^{(1)}} \ .
\ee
We find then that
\begin{eqnarray}
\lefteqn{\frac{\langle \psi(w_2, \bar w_2) \psi(w_1, \bar w_1) \rangle_n}{\langle \psi(w_2, \bar w_2) \psi(w_1, \bar w_1) \rangle_1} 
=}  \nonumber \\
&&
f \left(\zeta^{(n)}, \zeta^{(1)}, h \right)
f \left(\bar \zeta^{(n)}, \bar \zeta^{(1)}, \tilde h \right)
\ , 
\end{eqnarray}
where
\be
f(x, y, h) \equiv \frac{1}{n^{2h}} \left(\frac{x_1 x_2}{y_1 y_2} \right)^h \left( \frac{y_2-y_1}{x_2-x_1}\right)^{2h} \ .
\ee
If we take $t_2 = \infty$ and $t_1 = -\infty$, the expression simplifies,
\begin{eqnarray}
\frac{\langle \psi(\infty) \psi(-\infty) \rangle_n}{\langle \psi(\infty) \psi(-\infty) \rangle_1} 
&=& 
\frac{ 1}{n^{2\Delta}} \left( \frac{\sin\frac{\theta_2 - \theta_1}{2}}{\sin \frac{\theta_2 - \theta_1}{2n}}\right)^{2\Delta} 
\ .
\end{eqnarray}
The result (\ref{dSncorr}) then follows immediately and (\ref{dSEcorr}) by taking the $n \to 1$ limit of (\ref{dSncorr}).
The degeneracy factor $g$ comes from the fact that we can choose an orthonormal basis for operators $\psi_i$, $i=1, \ldots, g$ with dimension $\Delta$ where each $\psi_i$ contributes equally to (\ref{dSncorr}).  
A numerical confirmation of (\ref{dSEcorr}) is presented in figure \ref{EEplot} for free fermions.

We note in passing that a similar calculation was performed in ref.\  \cite{Alcaraz:2011tn}.  In that paper, the authors consider 
R\'enyi entropies in excited states of the CFT, for which they need to insert two operators $\psi(z, \bar z)$ on all of the $n$ sheets of the cylinder.  This makes the actual computation impractical for $n>2$. By contrast, the leading correction term in our calculation comes from inserting only two operators and may therefore be carried out for arbitrary $n$.

\vskip 0.1in

\noindent
{\it Limiting Forms for the Corrections:}
In the limit $\ell \ll L$, the correction (\ref{dSncorr}) can be written as a power series in $\ell/L$,
\begin{eqnarray}
\label{smallellSn}
\delta S_n &=& \left[ g \Delta \frac{1+n}{3n} \frac{\pi^2 \ell^2}{L^2}  + {\mathcal O}\left( \frac{\ell^4}{L^4} \right) \right] e^{-2 \pi \Delta \beta/L} \nonumber \\
&&
+ o(e^{-2 \pi \Delta \beta / L})
\ .
\end{eqnarray}
We can understand this form of the correction from a twist operator formulation of the R\'enyi entropy.  
Instead of working on the uniformized Riemann surface $\zeta^{(n)}$, we work on the individual cylinders coordinatized by $w$ but now with twist operators $P(\ell/2)$ and $P(-\ell/2)$ at the endpoints of the interval $A$.
We can write down a couple of terms in the OPE of the twist operators \cite{Headrick:2010zt,Calabrese:2009ez,Calabrese:2010he} \footnote{%
  We take there to be $2n$ twist operators, 2 on each cylindrical sheet of the Riemann surface.
  Note the coefficient of the stress tensor in the OPE and the exponent of the leading $\ell$ are fixed by the conformal dimension of 
  the twist operators, $h = \tilde h = (c/24)(1-1/n^2)$  \cite{Calabrese:2004eu}.
  Recall that the stress tensor $T(z)$ will appear with coefficient $2h/c$ in the OPE of two primary operators, provided that the  
  primaries are normalized such that the two point function takes the standard form (\ref{standardform}).  Here
  there is instead an overall multiplicative factor $c_n$.
  %$\langle \psi(w_1, \bar w_1) \psi(w_2, \bar w_2) \rangle = (w_1-w_2)^{-2h} (\bar w_1 - \bar w_2)^{-2 \tilde h}$.
%
}.
\begin{eqnarray}
\label{twistOPE}
\lefteqn{
P( \ell/2) \tilde P(- \ell/2) = c_n \ell^{-(c/6) (1-1/n^2)} \Biggl( 1 + \ldots  
}
\nonumber
\\
&& 
+ \frac{1}{12}\left(1 - \frac{1}{n^2} \right) \ell^2 (T(0) + \tilde T(0) )+ \ldots \Biggr) \ .
\end{eqnarray}
(A discussion of the normalization constants $c_n$
can be found in ref.\ \cite{Calabrese:2009qy}.)
The leading $(\ell/L)^2$ correction (\ref{smallellSn}) will then come from a three point function involving the stress tensor and the two quasi-primary fields $\lim_{t \to \infty} \psi(x,t)$ and $\lim_{t \to -\infty} \psi(x,t)$ producing, via the operator state correspondence, the first excited state in the Boltzmann sum (\ref{rhoexpand}).
Recall 
the three point function of the stress tensor with two quasi-primary fields $\psi(z, \bar z)$ of dimension $\Delta = h + \tilde h$ is
\be
\langle \psi(z_2, \bar z_2) T(z_3) \psi(z_1, \bar z_1) \rangle = h\frac{z_{12}^2}{z_{31}^2 z_{32}^2} \frac{1}{|z_{12}|^{2 \Delta} }\ .
\ee
If we now transform to the cylinder, $z = e^{2 \pi \i w / L}$, choosing $z_1 = 0$ and $z_2 = \infty$, we find that
\be
\frac{\langle  \psi(w_2, \bar w_2) T(w_3) \psi(w_1, \bar w_1) \rangle }{\langle \psi(w_2, \bar w_2) \psi(w_1, \bar w_1) \rangle } = -\frac{4 \pi^2}{L^2} h  + \frac{\pi^2 c}{6 L^2}\ .
\ee
The small $\ell$ correction should then be
\be
g \cdot  \frac{1}{1-n} \cdot n \cdot  \left(1 - \frac{1}{n^2} \right) \frac{\ell^2}{12} \cdot \left(  -\frac{4 \pi^2 \Delta}{L^2} \right) \cdot e^{-2 \pi \Delta \beta / L}\ ,
\ee
which matches with (\ref{smallellSn}).  The $g$ is the degeneracy factor.  The $1/(1-n)$ comes from the definition of the R\'enyi entropy.  The $n$ is the number of different sheets where we can compute this three point function.  The $(1-1/n^2)\ell^2/12$ comes from the OPE of the twist operators.  Finally, we get contributions from three point functions involving $T(w)$ and $\tilde T(\bar w)$ and a Boltzmann factor.  Note the Schwarzian derivative contribution to the three point function (24) does not contribute to (25) but instead
gets incorporated into the leading zero temperature part of the R\'enyi entropies.

In the limt $\ell \to L$, the correction (\ref{dSncorr}) becomes 
\begin{eqnarray}
\delta S_n &=&  \left[ -g\frac{n}{1-n} + {\mathcal O}\left( 1 - \frac{\ell}{L} \right)^{2 \Delta} \right] e^{-2 \pi \Delta \beta / L}   \nonumber \\
&&
+ o(e^{-2 \pi \Delta \beta/L} ) \ .
\end{eqnarray}
This expression traces its origin to the normalization of the thermal density matrix, i.e.\ the $e^{-2 \pi \Delta \beta / L}$ in the denominator of (\ref{rhoexpand}), and leads to a peculiar singularity in the limit $n \to 1$.

Interestingly, the analytic continuation $n \to 1$ is sensitive to the order of limits $\beta \to \infty$ and $\ell \to L$.  Above, we have taken $\beta \to \infty$ first.  If we instead first take $\ell \to L$, and then take $n \to 1$, the coefficient of $e^{-2 \pi \Delta \beta / L}$ now becomes singular in the limit $\beta \to \infty$.  
%Rather, we find the thermal entropy plus a universal temperature independent correction.  
Let us see how this result comes about in greater detail.  The $n$th R\'enyi entropy can be computed from the partition function of an $n$-sheeted copy of the torus, glued along the interval $0$ to $\ell$.  On each sheet of the Riemann surface, we insert the identity operator $\operatorname
{Id} = P P^{-1}$ along a spatial circle where $P$ shifts up a sheet and $P^{-1}$ shifts down.  Then, we move $P^{-1}$ so that it overlaps and partially cancels with the interval $0$ to $\ell$, leaving $P^{-1}$ along the interval $\ell$ to $L-\ell$.  
In the limit $\ell \to L$, we can replace the sewing prescription along the interval $\ell$ to $L-\ell$ with the OPE of twist operators that sit at $\ell$ and $L-\ell$.  The effect of the remaining $P^{-1}$ around the spatial circle is to glue the $n$ tori of length $\beta$ into a single torus of length $n \beta$.  
Using the OPE (\ref{twistOPE}), we find that
\begin{eqnarray}
\tr (\rho_A)^n &=& c_n^n (L-\ell)^{(-c/6)(n-1/n)}\frac{ \tr (\rho^n)}{(\tr \rho)^n} + \ldots 
%\nonumber \\
%&&
%{\mathcal O}((L-\ell)^2, (L-\ell)^{2 \Delta}) \Biggr)\ .
\label{trhoAn}
\end{eqnarray}
We will return to $L-\ell$ dependent corrections in a moment.
The leading $(L-\ell)$ dependence yields the universal single interval R\'enyi entropy for a CFT on a line while the remaining ratio of thermal density matrices gives the thermal R\'enyi entropy for the whole system.  The $n\to 1$ limit of the thermal R\'enyi entropy will yield the ordinary thermal entropy.  At low temperature for our system, the thermal R\'enyi entropy can be expanded,
\be
 \frac{ \tr (\rho^n)}{(\tr \rho)^n} = 
 \frac{1 + g e^{-2 \pi \Delta n \beta/L} + \ldots }{(1+ g e^{-2 \pi \Delta \beta / L} + \ldots)^n}   \ .
\ee
From the $n\to 1$ limit of (\ref{trhoAn}),
we may compute the entanglement entropy for the interval $A$:
\begin{eqnarray}
S_E &=& \Biggl[\frac{c}{3} \log \frac{L-\ell}{\epsilon} + g \left(1 + \frac{2 \pi \Delta \beta}{L} \right) e^{-2 \pi \Delta \beta / L} + \ldots \Biggr]  \nonumber
\\
&&
+ {\mathcal O}((L-\ell)^2,(L-\ell)^{2 \Delta}) \ .
\label{SEcorrtwo}
\end{eqnarray}
Note that the ${\mathcal O}(e^{-2 \pi \Delta \beta/L})$ correction matches well the result for free fermions, shown in the inset of figure \ref{EEplot}.

We expect an ${\mathcal O}(L-\ell)^2$ correction to (\ref{SEcorrtwo})  from expanding the universal leading $\log \sin (\pi \ell / L)$ piece of the entanglement and R\'enyi entropies on a circle.  In terms of the OPE (\ref{twistOPE}), this correction comes from the one point function of the stress tensor on a cylinder.  Note that this correction is temperature independent and does not come with a $e^{-2 \pi \Delta \beta / L}$ Boltzmann factor.   The leading correction that does come with a $e^{-2 \pi \Delta \beta / L}$ suppression should scale as ${\mathcal O}(L-\ell)^{2 \Delta}$.  This correction will come from a $\psi^2$ term in the OPE of the two twist operators \cite{Calabrese:2010he}.  By dimensional analysis, the coefficient of $\psi^2$ in the OPE contains the $(L-\ell)^{2\Delta}$ dependence while the Boltzmann suppression comes from two point functions of the $\psi$ operators separated by a distance $\beta$ on the torus of length $n \beta$ and width $L$.

\vskip 0.1in

\noindent
{\it Discussion:}
For pure states, the R\'enyi entropy of a region and its complement are equal, $S_n(A) = S_n(B)$, as follows from a Schmidt decomposition of the Hilbert space (see for example \cite{EislerPeschel}).  However, for thermal states, this symmetry is broken.  Indeed for our corrections (\ref{dSncorr}), $\delta S_n(\ell) \neq \delta S_n(L-\ell)$.    However, the R\'enyi entropies and our corrections 
do have the symmetry
$S_n(\ell) = S_n(n L - \ell)$.  To see this symmetry, imagine taking one of the twist operators around a spatial cycle on the torus $(n-1)$ times. Then, on the interval $(0,\ell)$ we have a permutation $P_n^n=1$, and on the complementary interval a permutation $P_n^{n-1}$ which is equivalent to $P_n$. (For the correction (\ref{dSncorr}), note that one should take the modulus of the ratio of sines.)

The corrections (\ref{dSncorr}) and (\ref{dSEcorr}) are very similar in spirit to universal corrections to two interval entanglement entropy on the plane \cite{Headrick:2010zt,Calabrese:2009ez,Calabrese:2010he}.  
For two intervals of length $\ell_1$ and $\ell_2$ whose centers are separated by a distance $r$, an OPE argument
predicts a leading correction to the R\'enyi entropies of the form
\be
\delta S_n = C_n \left( \frac{\ell_1 \ell_2}{n^2 r^2} \right)^{\Delta} \ ,
\ee 
where $\Delta$ is again the lowest conformal scaling dimension among the primary operators and $C_n$ is known and calculable.  

It would be interesting to see if similar formulae can be found for CFTs in higher dimension.

\vskip 0.1in

\noindent
{\it Acknowledgments:}
We would like to thank K.~Balasubramanian, P.~Calabrese, T.~Faulkner, T.~Hartman, S.~Hartnoll, M.~Headrick, M.~Kulaxizi, A.~Parnachev, T.~Nishioka, L.~Takhtajan, and E.~Tonni for discussion.
We thank the Kavli Institute for Theoretical Physics, Santa Barbara, where this research was carried out, for its hospitality.
This work was supported by the National Science Foundation (NSF) under Grant No.\ PHY11-25915.  J.~C. was 
supported in part by the Simons Foundation, and 
C.~H. by the NSF under Grants No.\ PHY08-44827 and PHY13-16617.  C.~H. thanks the Sloan Foundation for partial support.

\end{document}